\documentstyle[preprint,aps,psfig]{revtex}
\tighten
\begin{document}
\draft
\preprint{\vbox{Submitted to Physical Review B 
		          \hfill FSU-SCRI-98-81 \\
		          \null\hfill CU-NPL-1162 \\}}
\title{Mean-Field Theory for Spin Ladders \\  
       Using Angular-Momentum Coupled Bases}	
\author{J. Piekarewicz}
\address{Department of Physics and 
         Supercomputer Computations Research Institute, \\
         Florida State University, Tallahassee, FL 32306}
\author{J.R. Shepard}
\address{Department of Physics, \\
         University of Colorado, Boulder, CO 80309}
\date{\today}
\maketitle
 
\begin{abstract}
We study properties of two-leg Heisenberg spin ladders in a mean-field
approximation using a variety of angular-momentum coupled bases. The
mean-field theory proposed by Gopalan, Rice, and Sigrist, which uses a
rung basis, assumes that the mean-field ground state consists of a
condensate of spin-singlets along the rungs of the ladder. We
generalize this approach to larger angular-momentum coupled bases
which incorporate---by their mere definition---a substantial fraction
of the important short-range structure of these materials.  In these
bases the mean-field ground-state remains a condensate of spin
singlet---but now with each involving a larger fraction of the spins
in the ladder. As expected, the ``purity'' of the ground-state, as
judged by the condensate fraction, increases with the size of the
elementary block defining the basis. Moreover, the coupling to
quasiparticle excitations becomes weaker as the size of the elementary
block increases. Thus, the weak-coupling limit of the theory becomes
an accurate representation of the underlying mean-field dynamics. We
illustrate the method by computing static and dynamic properties of
two-leg ladders in the various angular-momentum coupled bases.
\end{abstract}

\narrowtext

\section{Introduction}
\label{sec:intro}

Spin ladders have received much recent attention due to their likely
relevance to high-$T_c$ superconductors.  The full arsenal of
theoretical many-body physics methods has been brought to bear on the
problem of spin ladders, including direct diagonalization
(DD)\cite{barn93}, Lanczos techniques\cite{dago94}, quantum Monte
Carlo (QMC)\cite{barn93}, density matrix renormalization group
(DMRG)\cite{whit94}, high-order perturbation theory augmented by
re-summation techniques\cite{weih97}, mappings to effective field
theories\cite{troy95,shel96} as well as quasi-analytic methods based
on mean-field approaches\cite{gopa94}. In the present work, we focus
on the last of these techniques and extend the mean-field theory (MFT)
of Gopalan, Rice and Sigrist\cite{gopa94}.  This extension includes a
derivation of generalized mean-field equations appropriate to any
angular-momentum coupled basis. In particular, we examine MFT's
utilizing bases whose elements are eigenstates of $2\!\times\!1$
(``rung''basis), $2\!\times\!2$ (``plaquette'' basis) and
$2\!\times\!4$ systems. The original work of Ref.~\cite{gopa94}
employed a rung basis. Indeed, the strong coupling limit for which
coupling within rungs is much greater than coupling between rungs---in
which case the ladder ground state can be well-approximated by a
configuration in which every rung is coupled to a spin singlet---was
the inspiration for their MFT in which rung singlets ``condense''. We
also do MFT calculations in the rung basis but obtain rather different
results from those of Ref.~\cite{gopa94}. The likely origin of these
differences is addressed below. We find our rung-basis MFT results in
much better agreement with known properties of 2-leg ladder---such as
the ground state energy per site, the spin gap, the spin correlation
length\cite{whit94}, the one-magnon dispersion relation\cite{piek98a},
and the strength of the $S(\pi,\pi)$ spin response\cite{piek98}---than
those of Ref.~\cite{gopa94}. Our results suggest that rung-basis MFT
is a relatively simple but still robust means of accounting for the
physics of spin ladders.

We also study plaquette- and $2\!\times\!4$-basis MFTs. In a previous
work\cite{piek97}, we showed that the plaquette basis offers some
significant advantages over simpler bases (including the rung basis)
when describing the physically relevant\cite{dago96} case of isotropic
spin ladders where intra-rung and inter-rung couplings are equal. We
find here that the plaquette basis yields an MFT which is superior to
that of the rung basis.  While the two sets of calculations give very
similar mean-field values for many ladder observables, we conclude
that only the plaquette basis results are reliable.  This
determination is based on the fact that our estimates of contributions
beyond mean-field theory are quite large for the rung basis but much
smaller for plaquettes. As our general approach holds for any
angular-momentum coupled basis, we also study the MFT utilizing a
$2\!\times\!4$ basis. Here, MFT results are comparable to those for
the rung and plaquette MFTs. We trace the lack of significant
improvement upon going to this rather complicated basis to truncations
which become increasingly severe as the basis is built from the
eigenstates of larger and larger systems.  In this context, the
plaquette basis MFT seems the best compromise between building enough
physics into the basis that MFT can account for the bulk of the
physics and, at the same time, avoiding severe truncations.

The paper is organized as follows: In Section~\ref{sec:formal}, we
present our angular-momentum coupled formalism for spin ladders in the
context of the rung basis. We also present our rung-basis MFT and
identify differences with Ref.\cite{gopa94}. In addition, we derive
weak-coupling expansions for the energy per site and spin gap which
are useful in understanding the physics embodied in the MFT equations
(and which additionally reproduce the full MFT results with surprising
accuracy, especially for the plaquette basis). We next indicate how to
make a perturbative estimate of corrections to the energy per site and
spin gap arising from terms in the Hamiltonian which are omitted at
the mean-field level. Finally we extend our MFT to
plaquette and $2\!\times\!4$ bases and indicate where and with what
justification truncations are made. Numerical results and discussion
of them appear in Section~\ref{sec:results}, while
Section~\ref{sec:conclusions} contains our conclusions.

\section{Formalism}
\label{sec:formal}

\subsection{Two-leg Ladder Hamiltonian}
\label{sec:twoleg}

 The Heisenberg Hamiltonian for a two-leg ladder with nearest-neighbor 
antiferromagnetic coupling is given by
\begin{equation}
    H=J_{\parallel}\sum_{\leftrightarrow}{\bf S}_{i}\cdot{\bf S}_{j}
     +J_{\perp}\sum_{\updownarrow}{\bf S}_{i}\cdot{\bf S}_{j} \;.
 \label{hrung} 
\end{equation}
Here $J_{\perp}$ and $J_{\parallel}$ are the strength of the
antiferromagnetic coupling along the rungs and chains of the
ladder, respectively~\cite{barn93}. In the strong-coupling 
limit, $\lambda\equiv~J_{\parallel}/J_{\perp}\rightarrow 0$, 
the interaction between rungs is very weak and the above 
separation of the Hamiltonian is, indeed, a natural one. In
this limit the Hamiltonian can be separated into a large
component $H_{0}$, which contains the strong intra-rung 
interactions, and a weak ``two-body'' potential $V$ 
between rungs. That is, $H=H_{0}+\lambda V$, where
\begin{eqnarray}
  \phantom{V}H_{0}&=&
  \sum_{r=1}^{N}{\bf S}_{1}(r)\cdot{\bf S}_{2}(r) \;, 
  \label{h0rung}                                  \\
  \phantom{H_{0}}V&=&\sum_{r=1}^{N}
       \left[{\bf S}_{1}(r)\cdot{\bf S}_{1}(r+1)
            +{\bf S}_{2}(r)\cdot{\bf S}_{2}(r+1)\right] \;.
 \label{vrung} 
\end{eqnarray} 
Note that we have set $J_{\perp}\equiv 1$ and that $r$ labels a 
specific rung in the $N$-rung ladder. Further, 
${\bf S}_{1}(r)[{\bf S}_{2}(r)]$ denotes the spin operator 
for the electron on the first[second] site of the $r$-th rung.
The rung basis is particularly suitable in the strong-coupling 
regime, as $H_{0}$ is diagonal in this basis. The eigenvalues and 
eigenvectors of $H_{0}$ are of the form:
\begin{equation}
   |\phi_{\ell m}\rangle = 
   |(s_{1}s_{2})\,\ell m\rangle \;; \quad
   \epsilon_{\ell}={1\over 2}\ell(\ell+1)-3/4 \;.
  \label{onerung} 
\end{equation}
Here the two spins along a definite rung in the ladder are coupled 
to a total angular momentum $\ell$, which can be zero or one, and 
projection $m$. Rung singlet and triplet creation operators are 
defined accordingly
\begin{equation}
    A^{\dagger}_{\ell m}(r) \equiv 
    \Big[
      c_{1}^{\dagger}(r)\otimes c_{2}^{\dagger}(r)  
    \Big]_{\ell m}\;,
  \label{defa}
\end{equation}
where $c^{\dagger}$ is an electron creation operator. Note that the 
rung singlet/triplet operators satisfy the usual commutation 
relations for bosons: 
\begin{equation}
    [A_{\ell m}(r),A^{\dagger}_{\ell' m'}(r')]=
    \delta_{\ell\ell'}\delta_{mm'}\delta_{rr'} \;.
 \label{comrel}
\end{equation}
We now proceed to rewrite the above Hamiltonian in a second-quantized
form. For the unperturbed part of the Hamiltonian, namely, the part 
that is diagonal in the rung basis, we obtain
\begin{equation}
    H_{0}=\sum_{r=1}^{N}\sum_{\ell=0}^{1}\hat{\ell}\epsilon_{\ell}
    \Big[
      A^{\dagger}_{\ell}(r)\otimes\widetilde{A}_{\ell}(r)
    \Big]_{0,0} = \sum_{r=1}^{N}
    \Big[
      \epsilon_{0}s^{\dagger}(r)s(r)-
      \epsilon_{1}{\bf t^{\dagger}}(r)\cdot{\bf \widetilde{t}(r)}
    \Big] \;,
 \label{hzerorspace}
\end{equation}
where we have defined $\hat{\ell}\equiv\sqrt{2\ell+1}$ and
\begin{equation}
      s^{\dagger}(r)     \equiv A^{\dagger}_{00}(r) \;; \quad
      t^{\dagger}_{m}(r) \equiv A^{\dagger}_{1m}(r) \;.
 \label{defst}
\end{equation}
Note that we have also introduced the ``tilde'' operators through the
definition: 
\begin{equation}
      \widetilde{t}_{m}=(-1)^{1-m}t_{-m} \;.
 \label{deftilde}
\end{equation}
As we have indicated explicitly in Eq.~(\ref{hzerorspace}) it is 
the operator ${\bf \widetilde{t}}$---not ${\bf t}$---that transforms
as a vector under rotations; the $m$-dependent phase factor in the 
above equation is necessary to enforce this condition. In analogy to 
the one-body term, the two-body potential energy [Eq.~(\ref{vrung})] 
can be written in a rotationally-invariant form
\begin{equation}
  V=\sum_{r=1}^{N}
   \sum_{\ell'_{1}\ell'_{2}\ell_{1}\ell_{2}j}\hat{\jmath}\,
   \langle\ell'_{1}\ell'_{2}j||V(1,2)||\ell_{1}\ell_{2}j\rangle
   \left[ 
   \Big(
     A^{\dagger}_{\ell'_{1}}(r)A^{\dagger}_{\ell_{2}'}(r+1)
   \Big)_{j}
   \Big(
     \widetilde{A}_{\ell_{1}}(r)\widetilde{A}_{\ell_{2}}(r+1)
   \Big)_{j}
   \right]_{00} \;.
 \label{secquanvee} 
\end{equation} 
Matrix elements of the potential have been computed in
Ref.~\cite{piek98a} and have been listed---along with the corresponding
matrix elements in other bases---in Table~\ref{tableone}. It is
important to mention that due to the symmetry of the potential only
two of these matrix elements are independent. We have chosen them to
be:
\begin{equation}
  \alpha\equiv \langle 011||V(1,2)||101\rangle 
  \quad {\rm and} \quad
  \beta \equiv \langle 110||V(1,2)||110\rangle \;.
 \label{defab}
\end{equation}

If we now assume periodic boundary conditions, then the total linear 
momentum becomes also a good quantum number. Thus, it is convenient 
to re-write the Hamiltonian in momentum space. To this end, we start
by performing a Fourier transformation of the spin singlet and triplet 
operators 
\begin{equation}
    A^{\dagger}_{\ell m}(r) = {1 \over \sqrt{N}} 
    \sum_{k} e^{ikr} A^{\dagger}_{\ell m}(k)\;;
    \qquad \left(k={2\pi\nu \over N}\;; \;\;\nu=1,2,\ldots,N \right) \;.
 \label{fourier}
\end{equation}
Moreover, we decomposed the Hamiltonian into three components:
\begin{equation}
  H = H_{0} + \lambda H_{1} + \lambda H_{2} \;.
 \label{h012} 
\end{equation}
The first term in this decomposition is simply the unperturbed
Hamiltonian of Eq.~(\ref{hzerorspace}) written in momentum space
\begin{equation}
   H_{0} = \sum_{k} 
      \Big[
       \epsilon_{0}s^{\dagger}(k)s(k)-
       \epsilon_{1}{\bf t^{\dagger}}(k)\cdot{\bf \widetilde{t}}(k) 
      \Big] \;.
 \label{hzerokspace}
\end{equation}
The second term represents that part of the two-body potential that
is bilinear in both the spin-singlet and spin-triplet operators. It
is given by
\begin{equation}
      H_{1} = \frac{1}{N}\sum_{k_{1}k_{2}q}\alpha\cos(q)
      \Big[
           {\bf t}^{\dagger}(k'_{1})\cdot{\bf t}^{\dagger}(k'_{2})
	   s(k_{2})s(k_{1}) -	
	   {\bf t^{\dagger}}(k'_{1})\cdot{\bf\widetilde{t}}(k_{2})
 	   s^{\dagger}(k'_{2})s(k_{1})+{\rm h.c.} 
      \Big] \;.
 \label{honekspace}
\end{equation}
Note that we have introduced the two ``final'' momenta by
defining  $k'_{1}\equiv k_{1}-q$ and $k'_{2}\equiv k_{2}+q$. 
Finally, $H_{2}$ is the remaining---quartic in the spin-triplet 
operator---part of the potential:
\begin{equation}
      H_{2} = \frac{1}{N}
      \sum_{j=0}^{2}
      \sum_{k_{1}k_{2}q}C_{j}\beta\cos(q)
      \left[
       \Big(
	{\bf t}^{\dagger}(k'_1){\bf t}^{\dagger}(k'_2)
       \Big)_{j}
       \Big(
        \widetilde{\bf t}(k_2)\widetilde{\bf t}(k_1)
       \Big)_{j}
      \right]_{00} \;,
 \label{htwokspace}
\end{equation}
where $C_{j}\equiv (-1)^{j+1}\hat{\jmath}\;(1-3\delta_{j0})/2$.
Note that there is no component of the potential having an odd 
number of spin-triplet operators because of the selection rule 
$\ell_{1}+\ell_{2}+\ell'_{1}+\ell'_{2}={\rm even}$~\cite{piek98a}.

We now conclude this section by introducing the ``grand-canonical'' 
Hamiltonian $K$ through the definition
\begin{equation}
  K = H - \mu N \;.
 \label{kamil}
\end{equation}
Here $\mu$ is the chemical potential and $N$ is the number operator 
given by
\begin{equation}
  N = \sum_{k}
      \Big[
       s^{\dagger}(k)s(k)+\sum_{m}
       {\bf t}^{\dagger}_{m}(k){\bf t}_{m}(k)
      \Big] = \sum_{k}
      \Big[
       s^{\dagger}(k)s(k)-
       {\bf t^{\dagger}}(k)\cdot{\bf \widetilde{t}}(k) 
      \Big] \;.
 \label{constraint}
\end{equation}
As long as no approximations are being made, the description of
the interacting system in term of either $H$ or $K$ is equally 
acceptable. The virtue of $K$, however, is in that it allows a
consistent treatment of the system even if the ``particle'' 
number is not conserved (i.e., if $[H,N]\ne 0$). Since the 
mean-field approximation of the next chapter is based on a 
Bogoliubov transformation, using the grand-canonical 
Hamiltonian $K$ provides a definite advantage over $H$. 

\subsection{Mean-Field Approximation}
\label{sec:mfa}
In this section we implement the mean-field approximation of Gopalan,
Rice, and Sigrist~\cite{gopa94}. The central dynamical assumption of
their model is that the spin-singlet bosons with zero linear momentum
form a condensate. Hence, it becomes natural to replace the
spin-singlet operators by ``c-numbers'', i.e.,
\begin{equation}
   s(k),s^{\dagger}(k) \longrightarrow 
   \sqrt{N}\bar{s}\,\delta_{k,0} \;.
 \label{sbosons}
\end{equation}
If this condition is now substituted on the above expressions for
$H_{0}$ and $H_{1}$ [Eqs.~(\ref{hzerokspace},\ref{honekspace}),
respectively] we obtain
\begin{mathletters}
 \begin{eqnarray}
   H_{0} &=& N\bar{s}^{2}\epsilon_{0} - \epsilon_{1}
    \sum_{k} {\bf t^{\dagger}}(k)\cdot{\bf \widetilde{t}}(k) \;, \\
   H_{1} &=& \bar{s}^{2}\sum_{k}\alpha\cos(k)
      \Big[
           {\bf t}^{\dagger}(k)\cdot{\bf t}^{\dagger}(-k) -
	   {\bf t}^{\dagger}(k)\cdot{\bf\widetilde{t}}(k) +
	   {\rm h.c.} 
      \Big] \;.
 \end{eqnarray}
 \label{hmf}
\end{mathletters}
These expressions are now bilinear in the spin-triplet operators and,
thus, amenable for diagonalization via a Bogoliubov transformation.
Note that $H_{2}$ [Eq.~(\ref{htwokspace})] remains quartic in the
spin-triplet operators. Hence, for the moment we will ignore this 
term in the determination of the mean-field ground state and will 
return to it later to compute its effect perturbatively. In this 
way, the mean-field Hamiltonian can be written in the following 
compact form
\begin{equation}
 K_{\rm MF}(\bar{s}^2,\mu)  = 
   N(\epsilon_{0}-\mu)\bar{s}^2    -
   \frac{3}{2}N(\epsilon_{1}-\mu)  +
   \frac{1}{2}\sum_{km}
   {\cal T}^{\dagger}_{m}(k)\,\Omega(k)\,{\cal T}_{m}(k) \;.
 \label{kmft}
\end{equation}
where
\begin{equation}
  {\cal T}_{m}(k) = 
   \left[
   \matrix{ {\bf t}_{m}(+k) \cr \cr
             \widetilde{\bf t}^{\dagger}_{m}(-k) }
   \right] \; \quad {\rm and} \quad
   \Omega(k)=
   \pmatrix{\phantom{-2}\Lambda(k) & -2\Delta(k)  \cr 
           -2\Delta(k) & \phantom{-2}\Lambda(k)} \;.
 \label{todef}
\end{equation}
Note that the two independent elements of the $2 \times 2$ matrix 
$\Omega(k)$ are given by
\begin{equation}
  \Lambda(k) = \epsilon_{1} - \mu + 2\lambda\alpha\bar{s}^2\cos(k)\; 
  \quad {\rm and} \quad
  \Delta(k)  = \lambda\alpha\bar{s}^2\cos(k) \;.
 \label{lamdel}
\end{equation}
The above mean-field Hamiltonian can now be diagonalized by means of
a standard Bogoliubov transformation. We obtain
\begin{equation}
 K_{\rm MF}(\bar{s}^2,\mu)  = 
   N(\epsilon_{0}-\mu)\bar{s}^2    -
   \frac{3}{2}N(\epsilon_{1}-\mu)  +
   \frac{3}{2}\sum_{k}\omega(k)    +
   \sum_{km}\omega(k)\gamma_{m}^{\dagger}(k)\gamma_{m}(k) \;,
 \label{kmftdiag}
\end{equation}
where 
\begin{equation}
\omega(k) = \sqrt{\Lambda(k)^{2}-4\Delta^{2}(k)} =
            (\epsilon_{1}-\mu)\sqrt{1+d\cos(k)} \;; \quad
            \Big(d\equiv4\lambda\alpha\bar{s}^{2}/
	    (\epsilon_{1}-\mu)\Big) \;.
 \label{omegak}
\end{equation}
Moreover, the new ``quasiparticle'' creation operators are defined by
\begin{equation}
  \gamma_{m}^{\dagger}(k) = u(k){\bf t}_{m}^{\dagger}(k)
                          - v(k)\widetilde{\bf t}_{m}(-k) \;,
 \label{quasi}
\end{equation}
where the coefficients generating the canonical 
transformation---satisfying $[u^{2}(k)\!-\!v^{2}(k)]\!=\!1$---are 
given by
\begin{equation}
    u(k)  = \sqrt{\frac{\Lambda(k)+\omega(k)}{2\omega(k)}} \;
             \quad {\rm and} \quad
    v(k)  = {\rm sgn}[\Delta(k)]
             \sqrt{\frac{\Lambda(k)-\omega(k)}{2\omega(k)}} \;.
\end{equation}
The mean-field ground state---the quasiparticle vacuum---is now 
determined by the condition
\begin{equation}
  \gamma_{m}(k)|\Phi_{0}\rangle=0\;, 
  \quad {\rm for \ all\ } k \ {\rm and}\ m\;.
 \label{qvacuum}
\end{equation}
As we have described in the appendix, we have carried out the
Bogoliubov transformation by diagonalizing an appropriate 
$2 \times 2$ random-phase-approximation (RPA) matrix. Although 
not explicitly treated in the appendix, the method has the virtue 
of being easily generalizable to matrices of arbitrary dimension.
We should mention that for the particular case of the rung basis 
(i.e., $\epsilon_{0}\!=\!-3/4,\;$ $\epsilon_{1}\!=\!1/4,\;$ and 
$\alpha\!=\!1/2$) our results are identical to those obtained in 
Ref.~\cite{gopa94}---with one exception. The two factors of $3/2$ 
appearing in Eq.~(\ref{kmftdiag}) appear as factors of
$1/2$ in Ref.~\cite{gopa94}. In our derivation these factors arise
from the zero-point motion of the three triplet ``oscillators'', 
one for every magnetic component of ${\bf t}$. We will examine the
quantitative impact of this difference in Sec.~\ref{sec:results}.

Ground-state observables can now be obtained by taking appropriate 
derivatives of the thermodynamic potential
\begin{equation}
  K_{0}(\bar{s}^2,\mu)  \equiv 
   \langle \Phi_{0}|K_{\rm MF}(\bar{s}^2,\mu)|\Phi_{0}\rangle =
    N(\epsilon_{0}-\mu)\bar{s}^2    -
    \frac{3}{2}N(\epsilon_{1}-\mu)  +
    \frac{3}{2}\sum_{k}\omega(k)    \;.
 \label{kgs}
\end{equation}
For example, the number of rungs can be recovered by using the
thermodynamic relation
\begin{equation}
   N = -\left(\frac{\partial K_{0}(\bar{s}^2,\mu)}{\partial\mu}\right)
   \;,
  \label{dkdmu}
\end{equation}
or equivalently, from the following transcendental equation 
\begin{equation}
   \bar{s}^{2}-\frac{5}{2}+\frac{3}{2}
   \Big[{\cal I}(d)-d{\cal I}'(d)\Big]=0
   \;; \quad
   {\cal I}(d)\equiv\int_{0}^{2\pi}\frac{dk}{2\pi}\sqrt{1+d\cos(k)} \;.
  \label{integral}
\end{equation}
This equation serves to write $d$---and, thus, the chemical 
potential $\mu$---in terms of $\bar{s}^{2}$. Note that the 
function $d(\bar{s}^{2})$ is ``universal''; it is independent 
of the value of the ratio of $J_{\parallel}$ to $J_{\perp}$ 
and of the choice of basis. Finally, the expectation value of 
the singlet condensate may be determined from minimizing the
ground-state energy with respect to $\bar{s}^2$
\begin{equation}
   \left(\frac{\partial E_{0}(\bar{s}^2)}{\partial\bar{s}^2}\right)=0\;;
   \quad 
   E_{0}(\bar{s}^2)\equiv 
   K_{0}(\bar{s}^2,\mu(\bar{s}^2))+\mu(\bar{s}^2)N\;.
  \label{deds2}
\end{equation}
Having determined the values of the chemical potential $\mu$ and the
singlet condensate $\bar{s}^2$, static observables---such as the
energy-per-site and the one-magnon dispersion relation 
$\omega(k)$---can be easily computed.

\subsection{Dynamic Spin Response}
\label{sec:response}
Perhaps more exciting is the possibility of computing the dynamic
spin response of the system. This kind of computation has proven to 
be particularly challenging in certain approaches---such as DMRG---as
transition matrix elements, in addition to the excitation spectrum,
must be computed. Yet the dynamic response constitutes one of the
most fundamental properties of the system and one that may soon be
extracted from new data on inelastic neutron scattering experiments
from single crystals.

For two leg-ladders one can define two independent spin responses, 
$S_{0}(q,\omega)$ and $S_\pi(q,\omega)$, 
according to the relative phase between the two spin operators 
along the rungs of the ladder. That is,
\begin{equation}
  S_{0,\pi}(q,\omega)=\sum_{n}
   \Big|\langle\Psi_{n}|S_{0,\pi}(q)|\Psi_{0}\rangle\Big|^{2}
   \delta(\omega-\omega_{n}) \;,
 \label{response}
\end{equation}
where $\Psi_{0}$ is the exact ground-state wavefunction of the 
system, $\Psi_{n}$ an excited state with excitation energy
$\omega_{n}$, and the transition operator is given by
\begin{equation}
  S_{0,\pi}(q)=\sum_{r} e^{iqr}
  \Big[{\bf S}_{1}(r)\pm{\bf S}_{2}(r)\Big]_z \;.
 \label{spops}
\end{equation}
Although similar in form, these two operators probe very different
aspects of the dynamics of the system. This can be seen most easily 
by expressing both transition operators in a second-quantized form:
\begin{mathletters}
 \begin{eqnarray}
  \Big[{\bf S}_{1}(r)+{\bf S}_{2}(r)\Big]_{m} &=& \sqrt{2}
  \Big[{\bf t}^{\dagger}(r)\otimes\tilde{\bf t}(r)\Big]_{1m} \;; \\
  \Big[{\bf S}_{1}(r)-{\bf S}_{2}(r)\Big]_{m} &=& 
  \Big[{\bf t}^{\dagger}_{m}(r)s(r)-
   s^{\dagger}(r)\tilde{\bf t}_{m}(r)\Big]\, 
   \smash{\mathop{\longrightarrow}\limits_{\rm MF}}\,
   \bar{s}\Big[{\bf t}^{\dagger}_{m}(r)-
         \tilde{\bf t}_{m}(r)\Big] \;.
 \label{s1pms2}
 \end{eqnarray}
\end{mathletters}
These equations imply that $S_{\pi}$ probes the low-energy part of 
the spectrum, as it couples the ground-state of the system to the 
one-magnon band. In contrast, $S_{0}$---a two-magnon operator---probes 
the high-energy part of the response. Thus, the $S_{\pi}$ response 
becomes instrumental in elucidating the important 
low-energy/low-temperature behavior of the system. 

Matrix elements of the above operators can be readily evaluated 
by expressing the spin-triplet operators in terms of the 
quasiparticle operators of Eq.~(\ref{quasi}). We obtain
\begin{mathletters}
 \begin{eqnarray}
  \frac{1}{N}S_{0}(q,\omega) &=& 2 
  \int_{0}^{2\pi}\frac{dk}{2\pi}
  \Big[u(k)v(k+q)-u(k+q)v(k)\Big]^{2}
  \delta\Big(\omega\!-\!\omega(k)\!-\!\omega(k+q)\Big) \;; 
  \label{s0ofq} \\
  \frac{1}{N}S_{\pi}(q,\omega) &=&
  \bar{s}^{2}\Big[u(q)-v(q)\Big]^{2}
  \delta\Big(\omega\!-\!\omega(q)\Big) = 
  \frac{\bar{s}^{2}}{\sqrt{1+d\cos(q)}} \,
  \delta\Big(\omega\!-\!\omega(q)\Big) \;.
  \label{spiofq}
 \end{eqnarray}
 \label{s0piofq}
\end{mathletters}
Note that, given $q$ and $\omega$, the delta function in
Eq.~(\ref{s0ofq}) constrains the integral over $k$ with the result that
not only is the $S_{0}$ response pushed to high-energy
but, in addition, it is strongly fragmented.  This is in contrast to
the $S_{\pi}$ response that is predicted to be concentrated in a
single excitation at $\omega\!=\!\omega(q)$. At least for the
$S_{\pi}(q\!=\!\pi,\omega)$ response, this is in good agreement with
an earlier calculation~\cite{piek98} on a $2\times 16$ ladder that
predicts almost 95 percent of the strength to be concentrated at an
excitation energy equal to the singlet-triplet gap. Finally, having 
computed the two dynamic response functions the spin-spin correlation 
function per site can be easily extracted:
\begin{equation}
  \langle\Phi_{0}|S_{z}(0)S_{z}(r)|\Phi_{0}\rangle =
  \frac{1}{8} \int_{0}^{2\pi}\frac{dq}{2\pi}e^{iqr}
  \Big[S_{0}(q)+S_{\pi}(q)\Big]/N \;, 
 \label{spspin}
\end{equation}
where the static structure factors have been defined by
\begin{equation}
  S_{0,\pi}(q) = \int_{0}^{\infty} d\omega\, S_{0,\pi}(q,\omega) \;.
 \label{ssfactor}
\end{equation}
Note that in Eq.~(\ref{spspin}) both spins are located 
on the same chain of the ladder at a distance of $r$ rungs.
As we will see shortly the long-range behavior of the 
spin-spin correlation function is dominated by the static 
structure factor $S_{\pi}(q)$, namely,
\begin{equation}
  \langle\Phi_{0}|S_{z}(0)S_{z}(r)|\Phi_{0}\rangle \approx
  \frac{1}{8} \int_{0}^{2\pi}\frac{dq}{2\pi}e^{iqr}
  S_{\pi}(q)/N =
  \frac{\bar{s}^{2}}{8} \int_{0}^{2\pi}\frac{dq}{2\pi}
  \frac{e^{iqr}}{\sqrt{1+d\cos(q)}} \;.
 \label{spspin2}
\end{equation}
We can see that, in the spirit of the stationary-phase approximation,
the principal contribution to the integral at large $r$ occurs where 
the rapid oscillation of the exponential factor is canceled by rapid 
changes in $[1+d\cos(q)]^{-1/2}$. This will happen near any of the 
singularities of the latter factor. Indeed, upon using the
steepest-descent technique we obtain the expected exponential behavior 
of the spin-spin correlation function and our analytic approximation 
for the correlation length:
\begin{equation}
  \langle\Phi_{0}|S_{z}(0)S_{z}(r)|\Phi_{0}\rangle \sim
  e^{-r/\xi}\;; \quad
  \xi=1/\ln\left[d/(1-\sqrt{1-d^2})\right] . 
 \label{spspin3}
\end{equation}

\subsection{Weak-Coupling Expansions}
\label{sec:weak}
To gain further insight into the mean-field theory of
Ref.~\cite{gopa94} we derive in this section weak-coupling
($\lambda\!\ll\!1$) expansions for various observables.  We start this
procedure with an expansion of Eq.~(\ref{integral}), which determines
the parameter $d$---or equivalently the chemical potential---in terms
of the parameter $\bar{t}^{2}\!\equiv\!1\!-\!\bar{s}^2$.  That is,
\begin{equation}
  \left(\frac{d^2}{16}\right) +
  \frac{45}{4}\left(\frac{d^2}{16}\right)^{2} +
  {\cal O}(d^{6}) =
  \frac{2}{3}\bar{t}^{2} \; \longrightarrow \;
  \left(\frac{d^2}{16}\right) =
  \frac{2}{3}\bar{t}^{2} - 5\bar{t}^{4} + {\cal O}(t^{6}) \;.
 \label{doft2}
\end{equation}
This equation can now be substituted into the ground-state energy
to yield an expression that is correct to fourth-order in the small
parameter $\bar{t}={\cal O}(\lambda)$:
\begin{equation}
   E_{0}(\bar{t}) = \epsilon_{0} - \delta \bar{t} 
                  + \Delta_{0}\bar{t}^{2}
                  - \frac{\delta}{4}\bar{t}^{3} 
                  + {\cal O}(\bar{t}^{5}) \;; 
                  \quad {\rm where} \;
                  \delta\equiv\sqrt{6}\lambda|\alpha| \;.
 \label{eoft}
\end{equation}
By minimizing the above expression for the energy with respect to 
$\bar{t}$, ground-state---and even some excited-state---observables 
can be determined. Indeed, from the minimization procedure the singlet 
condensate is readily obtained
\begin{equation}
  \bar{s}^{2} = 1 - \frac{3}{2} 
    \left(
     \frac{\lambda\alpha}{\Delta_{0}}
    \right)^{2}
    \left[
      1 + \frac{9}{4} 
     \left(
      \frac{\lambda\alpha}{\Delta_{0}}
     \right)^{2} + {\cal O}(\lambda^{4})
     \right] \;.
 \label{sbarsquare}
\end{equation}
Moreover, having determined the behavior of the singlet condensate 
in terms of $\lambda$, one can generate---through Eqs.~(\ref{omegak}) 
and (\ref{doft2})---the corresponding behavior of the chemical
potential. This is all that is required to compute the mean-field
observables. For example, we obtain the following perturbative 
expansions for the ground-state energy per rung and for the 
singlet-triplet gap, respectively:
\begin{mathletters} 
\begin{eqnarray}
    E_{0} &=& \epsilon_{0} - \frac{3}{2}
     \frac{(\lambda\alpha)^{2}}{\Delta_{0}}
    \left[
      1 + \frac{3}{4} 
     \left(
      \frac{\lambda\alpha}{\Delta_{0}}
     \right)^{2} + {\cal O}(\lambda^{4})
     \right] \;,
     \label{perte} \\ 
     \Delta &=& \Delta_{0} - 2\lambda|\alpha|
     \left[1 - 
      \frac{1}{2}
      \left(
       \frac{\lambda|\alpha|}{\Delta_{0}}
      \right) +
      \frac{1}{2}
      \left(
       \frac{\lambda|\alpha|}{\Delta_{0}}
      \right)^{2} + {\cal O}(\lambda^{3}) 
      \right] \;.
      \label{pertgap}
 \end{eqnarray}
 \label{observables}
\end{mathletters}
Note that the small parameter in the above expansions 
is given by $\lambda\alpha/\Delta_{0}$.

\subsection{Perturbative Contribution From $H_{2}$}
\label{sec:pertur}
An important assumption of the mean-field approximation of
Ref.~\cite{gopa94} is that $H_{2}$---the term of the Hamiltonian
quartic in the spin-triplet operators
[Eq.~(\ref{htwokspace})]---changes the mean-field results only
slightly. To test the validity of this assumption we now estimate
the effect of $H_{2}$ on the ground-state energy and on the 
one-magnon dispersion relation using first-order perturbation 
theory. 

The first-order correction to the ground-state energy 
due to $H_{2}$ is given by
\begin{equation}
  \Delta E_{0} = \lambda \langle \Phi_{0}|H_{2}|\Phi_{0}\rangle \;.
 \label{ecorrection}
\end{equation}
To evaluate this correction to the energy the four spin-triplet 
operators in $H_{2}$ [Eq.~(\ref{htwokspace})] are expanded in 
terms of the quasiparticle operators---$\gamma_{m}(k)$ and 
$\gamma^{\dagger}_{m}(k)$---and the resulting matrix elements 
are evaluated using Wick's theorem. We obtain,
\begin{equation}
  \Delta E_{0}/N = 3\lambda\beta
  \left[
   \Big( 
    \int_{0}^{2\pi}\frac{dk}{2\pi}u(k)\cos(k)v(k)
   \Big)^{2} -
   \Big( 
    \int_{0}^{2\pi}\frac{dk}{2\pi}v(k)\cos(k)v(k)
   \Big)^{2}
   \right] \;.
 \label{deltae0}
\end{equation}
Note that only those combinations of quasiparticle operators that 
conserved quasiparticle number need to be considered. The 
corresponding perturbative contribution to the one-magnon band is 
slightly more complicated to evaluate. Still, it can be efficiently 
computed from the following double-commutator formula:
\begin{equation}
  \Delta\omega(k) = \lambda
  \Big[ 
   \langle \Phi_{0}|\gamma_{m}(k)\,H_{2}\,
   \gamma_{m}^{\dagger}(k)|\Phi_{0}\rangle -
   \langle \Phi_{0}|H_{2}|\Phi_{0}\rangle  
  \Big] =
  \lambda\langle\Phi_{0}| 
  \Big[
    \gamma_{m}(k),[H_{2},\gamma_{m}^{\dagger}(k)]
  \Big]|\Phi_{0}\rangle \;.
 \label{gapcorrection}
\end{equation}
In this way the resulting expression for the first-order correction 
to the dispersion relation becomes
\begin{eqnarray}
  \Delta \omega(k) = 2\lambda\beta \Bigg(
     2u(k)v(k)\cos(k)
    \int_{0}^{2\pi}\frac{dq}{2\pi}u(q)\cos(q)v(q) && \nonumber \\
     - \Big[u^{2}(k)+v^{2}(k)\Big]\cos(k)
    \int_{0}^{2\pi}\frac{dq}{2\pi}v(q)\cos(q)v(q) && \Bigg)  \;.
 \label{deltaomega}
\end{eqnarray}

\subsection{Mean-Field Theory with Improved Bases}
\label{sec:bases}
The mean-field theory of the previous section is fully characterized
by three parameters: the spin-singlet energy $\epsilon_{0}$, the
spin-triplet energy $\epsilon_{1}$, and the singlet-triplet matrix
element $\alpha$ (if corrections to the mean-field observables need to
be computed, an additional triplet-triplet matrix element, $\beta$, is
necessary). This suggests that the generalization of the mean-field
method of Gopalan, Rice, and Sigrist~\cite{gopa94} to larger
angular-momentum coupled bases is straightforward. Moreover, the
advantage of the larger bases is that they incorporate part of the
short-range structure of the system---and, thus, an important part of
the physics---exactly.

One such basis is the ``plaquette'' basis introduced by us in 
Ref.~\cite{piek97}. The plaquette basis is generated from  
coupling the four spins in a $2 \times 2$ lattice. Specifically, 
the two diagonal pairs of spins are coupled to well-defined total 
angular momenta, $\ell_{14}$ and $\ell_{23}$, which can equal 
zero or one. These two ``link'' angular momenta are in turn
coupled to a total plaquette angular momentum $j$ and projection 
$m$. The physically appealing feature of the plaquette basis is 
that the $2 \times 2$ ladder Hamiltonian with isotropic 
$(J_{\parallel}=J_{\perp}\equiv 1)$ coupling is diagonal in the 
plaquette basis. That is
\begin{equation}
  H(2\!\times\!2)\Big|\ell_{14}\ell_{23},jm\Big\rangle = {1 \over 2}
  \Big[
      j(j+1)-\ell_{14}(\ell_{14}+1)-\ell_{23}(\ell_{23}+1)
  \Big]
  \Big|\ell_{14}\ell_{23},jm\Big\rangle \;. 
 \label{etwosite}
\end{equation}
The lowest spin-singlet $(j\!=\!0)$ eigenstate arise from coupling the
two link angular momenta to their maximum value of
$\ell_{14}\!=\!\ell_{23}\!=\!1$. The corresponding value for the
energy is given by $\epsilon_{0}\!=\!-2$. Next comes the three-fold
degenerate spin-triplet $(\ell_{14}\!=\!\ell_{23}\!=j\!=\!1)$
eigenstate with an energy of $\epsilon_{1}\!=\!-1$. Note that these
four states---out of a total of sixteen---define the low-energy basis
the we have used earlier in a study of the dynamic
spin response~\cite{piek98}. It is precisely these four low-energy
states that we retain here to generalize the mean-field theory of
Ref.~\cite{gopa94} to the plaquette basis. The sole remaining
parameter needed to define the mean-field theory is the 
singlet-triplet matrix element $\alpha$. In the plaquette basis it 
is given by $\alpha\!=\!-1/3$. It is interesting to note that the
ratio of $\alpha$ to the unperturbed gap $\Delta_{0}$---which is a 
measure of the ``goodness'' of the basis 
[see Eq.~(\ref{observables})]---improves as one goes 
from the rung to the plaquette basis: $|\alpha/\Delta_{0}|$ goes 
from $1/2$ to $1/3$.

We can also compute the two dynamic spin responses in the plaquette
basis. Indeed, the $S_{0}$ and $S_{\pi}$ operators of
Eq.~(\ref{spops}) can be re-expressed in terms of the four spin 
operators along a $2\!\times\!2$ plaquette. We obtain
\begin{mathletters}
 \begin{eqnarray}
  S_{0}(q) &=& e^{-iq/2}\sum_{p} e^{i2qp}
   \Bigg[ 
   \cos(q/2)
   \Big(
    {\bf S}_{1}(p)+{\bf S}_{2}(p)+{\bf S}_{3}(p)+{\bf S}_{4}(p)
   \Big)_{z}  \nonumber \\ 
    && \hskip 2.15cm
   -i\sin(q/2)
   \Big(
    {\bf S}_{1}(p)+{\bf S}_{2}(p)-{\bf S}_{3}(p)-{\bf S}_{4}(p)
   \Big)_{z}
   \Bigg] \;; \\
  S_{\pi}(q) &=& e^{-iq/2}\sum_{p} e^{i2qp}
   \Bigg[ 
   \cos(q/2)
   \Big(
    {\bf S}_{1}(p)-{\bf S}_{2}(p)+{\bf S}_{3}(p)-{\bf S}_{4}(p)
   \Big)_{z}  \nonumber \\ 
    && \hskip 2.15cm
   -i\sin(q/2)
   \Big(
    {\bf S}_{1}(p)-{\bf S}_{2}(p)-{\bf S}_{3}(p)+{\bf S}_{4}(p)
   \Big)_{z}
   \Bigg] \;.
 \end{eqnarray}
\end{mathletters}
Note that ${\bf S}_{1}(p)$$[{\bf S}_{3}(p)]$ and 
          ${\bf S}_{2}(p)$$[{\bf S}_{4}(p)]$ denote the first and
second spin operators along the first[second] rung of the $p$-th 
plaquette. Moreover, the sum over $p$ now runs over all plaquettes 
in the ladder and the Fourier sum includes $2q$---not $q$---as the 
variable conjugate to $p$. In spite of their apparent 
complexity the above expressions are identical to the ones given 
in Eq.~(\ref{spops}). The virtue of writing them in such a form
becomes evident as one takes their mean-field limit. Indeed, in 
this limit the dynamic spin responses take the following simple form:
\begin{mathletters}
 \begin{eqnarray}
  \frac{1}{N}S_{0}(q,\omega)   &=& 0 \;; \\ 
  \frac{1}{N}S_{\pi}(q,\omega) &=&
  \frac{8}{3}\bar{s}^{2}
  \frac{\sin^{2}(q/2)}{\sqrt{1+d\cos(2q)}} \,
  \delta\Big(\omega\!-\!\omega(2q)\Big) \;.
 \end{eqnarray}
 \label{s0piplaq}
\end{mathletters}
Perhaps more than anything else these expressions capture the essence
of the mean-field approximation in the plaquette basis. Clearly, the
$S_{0}$ response is not zero [see. Eq.~(\ref{s0ofq})]. Yet, it is
small and its strength is fragmented and located at very high 
excitation energy. The ``low-energy'' plaquette basis---the truncated 
basis obtained by retaining only the lowest spin-singlet and spin-triplet
states---can not account for the physics at high excitation energy. In
contrast the $S_{\pi}$ response, which couples the ground-state to
the one-magnon band, probes the low-excitation part of the spectrum
and should be well described using the low-energy plaquette basis.

One could continue to systematically improve the basis. For example,
the spectrum of isotropic $2\times 4$ and $2\times 8$ ladders have
been previously calculated in Ref.~\cite{piek98}. In particular, the
lowest singlet and triplet energies of the $2\times 4$ ladder (with
open boundary conditions) are given by $\epsilon_{0}\!=\!-4.29307$ and
$\epsilon_{1}\!=\!-3.52286$, respectively. Moreover, from the $2\times
8$ calculation the singlet-triplet matrix element can also be
extracted; it is given by: $\alpha\!=\!-0.20805\!\simeq\!-1/5$.  Thus,
the mean-field theory can be further extended to the $2\times 4$
basis. Note that for this basis
$|\alpha/\Delta_{0}|\!\sim\!1/4$. Although one has obtained a better
expansion parameter in this basis relative to the plaquette basis, the
improvement has been somewhat reduced by the fact that as the
singlet-triplet matrix element $\alpha$ goes down so does the
unperturbed gap. Moreover, the truncation errors increase
considerably; while one retains the same four states in both bases,
the dimension of the vector space goes from 16 in the plaquette basis
to 256 in the $2\times 4$ basis. This will become an important issue
in our final analysis. Values for the mean-field parameters in the 
various bases are tabulated in Table~\ref{tableone}.

\section{Results}
\label{sec:results}
In this section we report mean-field results for the energy-per-site, 
the singlet-triplet gap, the one-magnon dispersion relation, the
spin-spin correlation function, and the dynamic spin response using 
various angular-momentum coupled bases. Because of the nature of 
these bases, any comparison  among them will only be carried out at 
the isotropic $(\lambda\!=\!1)$ point.

\subsection{Rung Basis}
\label{sec:resrung}

\begin{figure}
\centerline{
  \psfig{figure=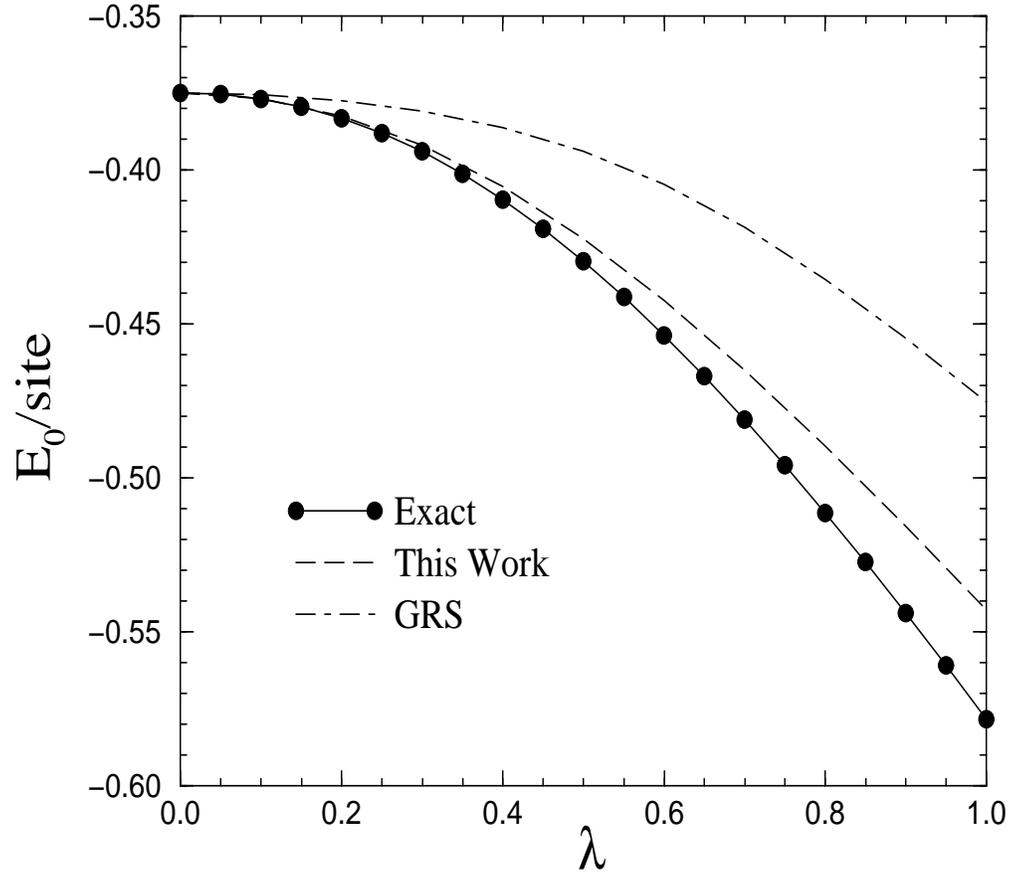,height=4.8in,width=5.2in,angle=0}}
 \vskip 0.1in
 \caption{Ground-state energy-per-site (in units of $J_{\perp}$)
          as a function of $\lambda$. The filled circles are 
          ``exact'' results obtained from a very high-order
          perturbative expansion~\protect\cite{weih97}, while 
	  GRS are the results of Gopalan, Rice, and 
	  Sigrist~\protect\cite{gopa94}.}
 \label{figone}
\end{figure}

We start with a presentation of our results in the rung basis.  In
Table~\ref{tabletwo} we have listed the ground-state energy-per-site
for three values of $\lambda$. The second column contains the
mean-field numbers reported in Ref.~\cite{gopa94}, while the third
column contains the corresponding values as obtained by us. The
difference between these values originates solely from the treatment
of the zero-point motion; recall that a factor of $1/2$, rather than
$3/2$, was used in Ref.~\cite{gopa94}. While the impact of this error
on the spin gap has been discussed recently in Ref.~\cite{norm96}, we
feel compelled to address it here as well, as the mean-field theory of
Gopalan, Rice, and Sigrist seems to work much better for most
observables---not only for the spin gap---than their original results
might suggest. Indeed, even for the most unfavorable case of
$\lambda\!=\!1$, the corrected mean-field results are within 6
percent---not 18 percent---of the exact answer.  Fig.~\ref{figone}
presents these results in graphical form and serves as a further
testimony to this fact.

\begin{figure}
\centerline{
  \psfig{figure=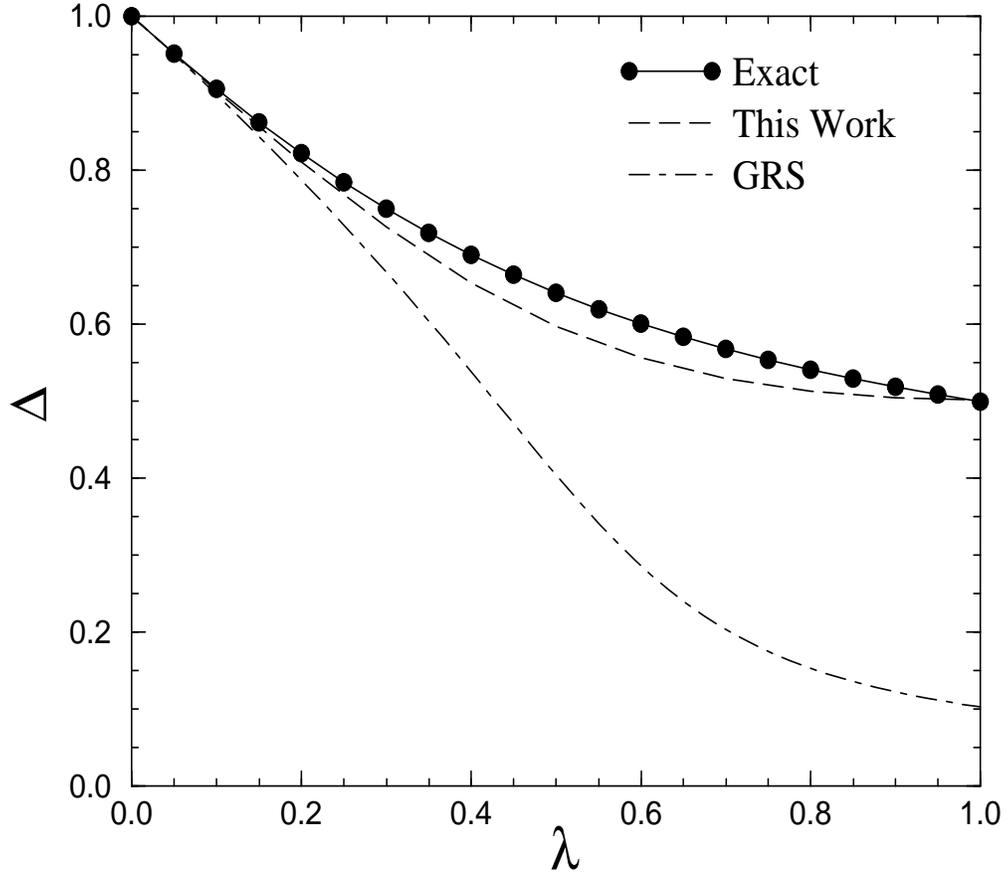,height=4.8in,width=5.2in,angle=0}}
 \vskip 0.1in
 \caption{Singlet-triplet gap (in units of $J_{\perp}$)
          as a function of $\lambda$. The filled circles
          are ``exact'' results obtained from a very high-order
          perturbative expansion~\protect\cite{weih97}, while 
          GRS are the results of Gopalan, Rice, and 
	  Sigrist~\protect\cite{gopa94}.}
 \label{figtwo}
\end{figure}

For the spin gap the improvement is even more dramatic. In
Table~\ref{tabletwo} and Fig.~\ref{figtwo} we display results for the
singlet-triplet gap (in units of $J_{\perp}$) as a function of
$\lambda$. The filled circles in the figure (labeled ``exact'') were
obtained from a very high-order perturbative expansion~\cite{weih97}
supplemented by a Pad\'e analysis. This plot shows conclusively that
the mean-field theory of Ref.~\cite{gopa94} captures an important part
of the physics governing the ladder materials. While the perfect
agreement at the isotropic point seems a bit fortuitous, it is clear
that the mean-field theory provides---in addition to an intuitive
physical picture---a very robust starting point for more sophisticated
calculations.

\begin{figure}
\centerline{
  \psfig{figure=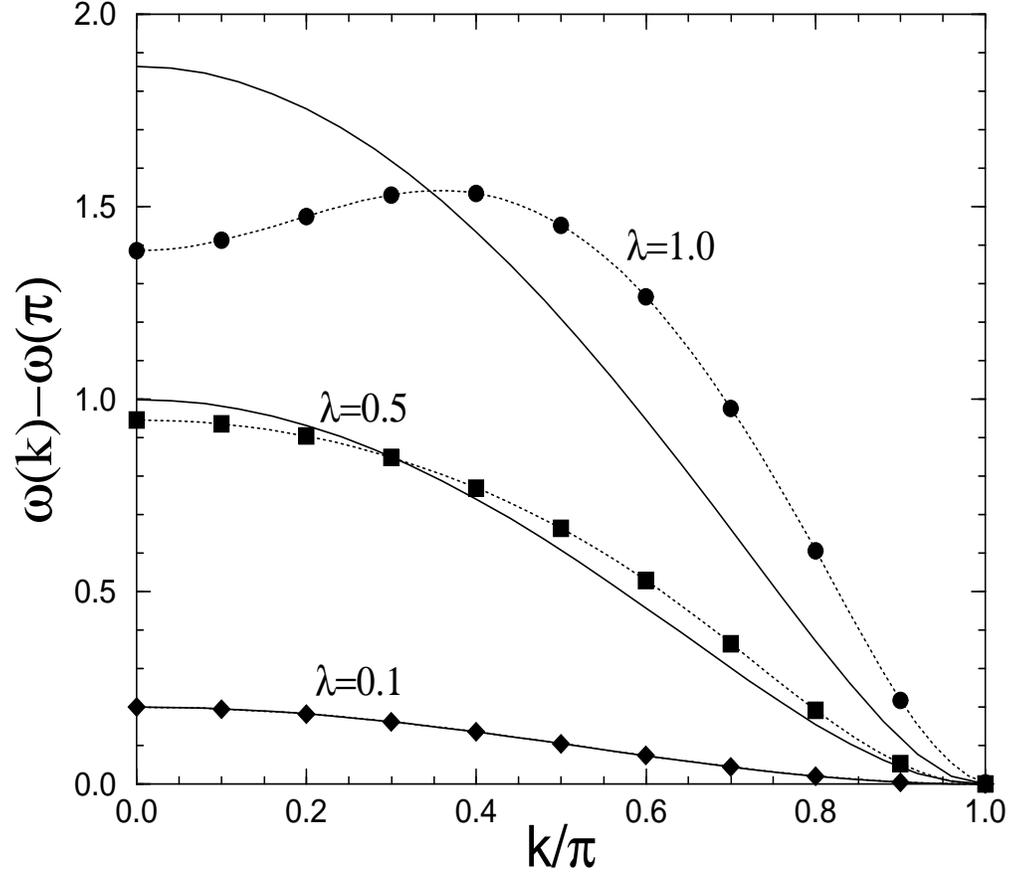,height=4.8in,width=5.2in,angle=0}}
 \vskip 0.1in
 \caption{Spin-triplet dispersion band relative to the band minimum 
          (in units of $J_{\perp}$) as a function of momentum.
          The filled symbols are exact 
	  results~\protect\cite{piek98a,barnes94} (see text); the dotted lines 
	  joining them were obtained from a cubic-spline fit.} 
 \label{figthree}
\end{figure}

In Fig.~\ref{figthree} we show our mean-field results for the
spin-triplet dispersion band (in units of $J_{\perp}$) relative to the 
value at the band minimum $\Delta=\omega(\pi)$. The filled circles for 
$\lambda\!=\!0.1$ and $0.5$ are the result of a perturbative 
calculation~\cite{piek98a}, while those at $\lambda\!=\!1$ were obtained
from the fit to Lanczos results carried out in Ref.~\cite{barnes94}. 
We observe that for $\lambda\!\leq\!0.5$ the agreement with exact 
results is relatively good even for the states at the top of the 
one-magnon band.

\begin{figure}
\centerline{
  \psfig{figure=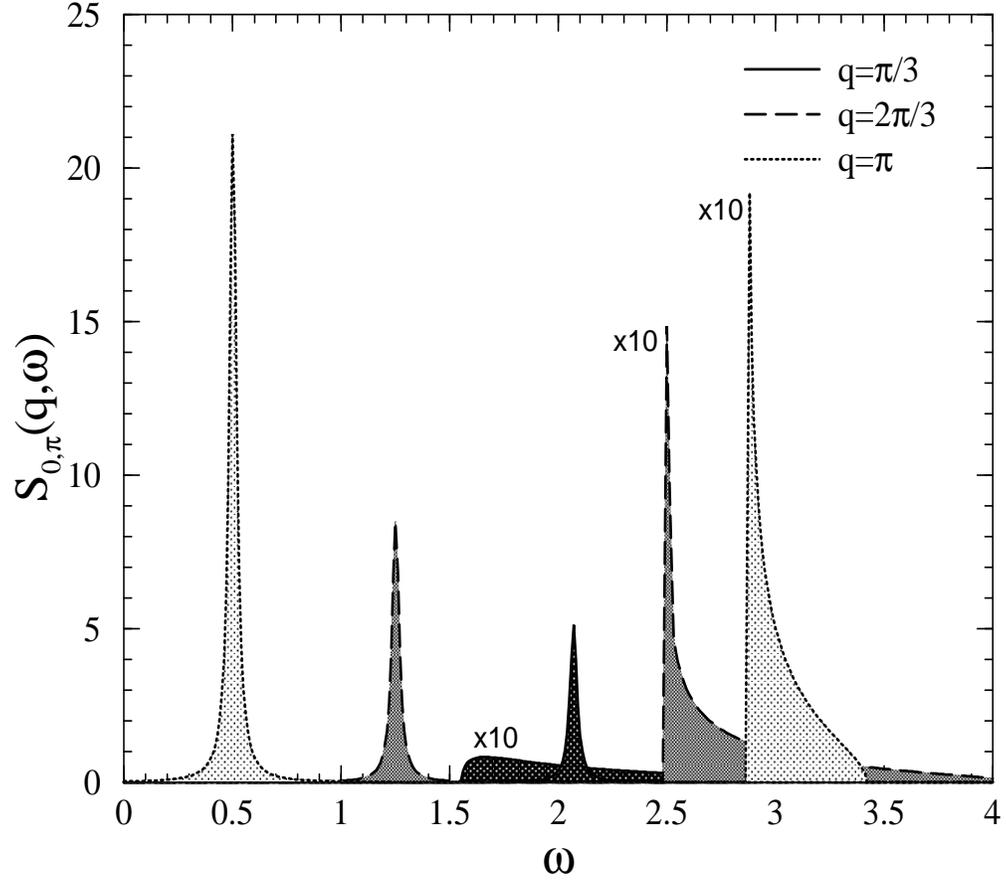,height=4.8in,width=5.2in,angle=0}}
 \vskip 0.1in
 \caption{Dynamic spin responses $S_{0}$ and $S_{\pi}$ in the rung
          basis as a function of $\omega$ for various values of $q$.  
	  The responses were computed at the isotropic value of 
	  $\lambda\!=\!1$. All $S_{0}$ responses have been multiplied 
	  by a factor of 10, while the $S_{\pi}$ responses include 
          a smearing factor of 0.02.}
 \label{figfour}
\end{figure}

\begin{figure}
\centerline{
  \psfig{figure=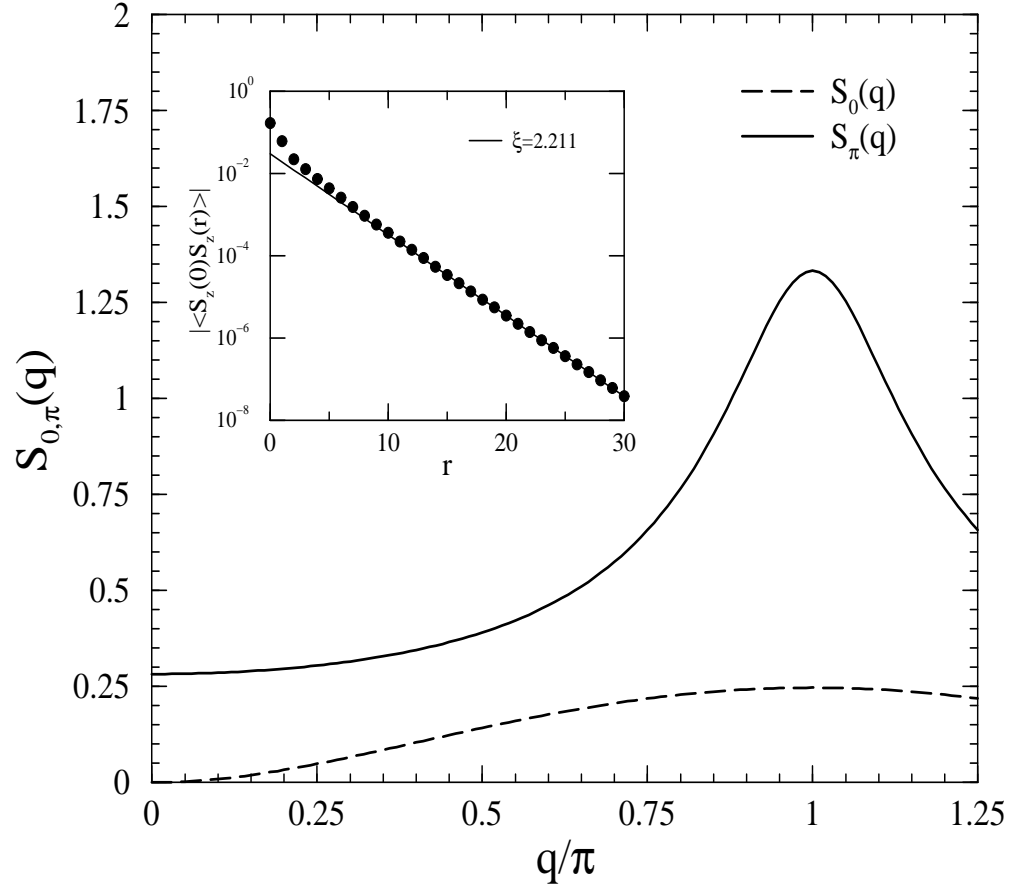,height=4.8in,width=5.2in,angle=0}}
 \vskip 0.1in
 \caption{Static structure factors $S_{0}$ and $S_{\pi}$ in
          the rung basis as a function of $q$ for $\lambda\!=\!1$. 
	  The inset shows a logarithmic plot of the spin-spin 
	  correlation as a function of r. A value of 
          $\xi\!=\!2.211$ is obtained for the spin-spin
          correlation length from the slope of the straight-line fit.} 
 \label{figfive}
\end{figure}

We now proceed to discuss the two dynamic spin responses. In 
Fig.~\ref{figfour} we display our results for $S_{0}(q,\omega)$
and for $S_{\pi}(q,\omega)$ as a function of the excitation energy
$\omega$ for various values of the momentum transfer $q$. For clarity,
all of the $S_{0}$ responses have been multiplied by a factor of ten,
while the $S_{\pi}$ responses include an artificial smearing of the
delta function:
\begin{equation}
  \delta(\omega\!-\!\omega_{0}) \approx
  \frac{\eta/\pi}{(\omega\!-\!\omega_{0})^{2}+\eta^2}\;; 
  \quad \eta=0.02 \;.
  \label{deltafcn}
\end{equation}
As suggested in our earlier discussion the $S_{0}$ response is indeed
small and---as it probes the two-magnon structure of the system---it
is located at high-excitation energy and its strength is strongly
fragmented. In contrast, the $S_{\pi}$ response is concentrated in a
single excitation and, at least for $q\!\simeq\!\pi$, it is located at
low-excitation energy $(\omega\gtrsim\Delta)$. The fact that all the
$S_{\pi}$ strength is concentrated in one fragment suggests that, once
the dispersion band is computed [see. Fig.~\ref{figthree}], all that
remains to fully characterize the $S_{\pi}(q,\omega)$ response is to
determine the static structure factor $S_{\pi}(q)$.  Such a plot is 
displayed in Fig.~\ref{figfive} where $S_{0}(q)$ also appears. The figure
clearly shows that $S_{0}(q)$ is 
not only appreciably smaller than $S_{\pi}(q)$ but it is also considerably
smoother. This suggests that the large Fourier components of the sum
of the two structure factors---and thus the long-range behavior of the
spin-spin correlation function---will, indeed, be dominated by 
$S_{\pi}(q)$. Thus, Eq.~(\ref{spspin3}) can be used to estimate 
the spin-spin correlation length:
\begin{equation}
   \xi\approx \frac{1}{\ln\left[d/(1-\sqrt{1-d^2})\right]}  
   \smash{\mathop{\longrightarrow}\limits_{d=0.914}}=2.323\;.
 \label{xirung}
\end{equation}
Note that a value of $\xi=2.211$ for the correlation length, extracted 
from an exponential fit to the full numerical evaluation of the 
spin-spin correlation function, compares well with the simple analytic
estimate given above---but differs substantially from the exact value 
of $\xi=3.19$~\cite{whit94}.

\subsection{Extension to Larger Bases}
\label{sec:extension}
In this section we compare mean-field results obtained on a variety of
angular-momentum coupled bases. We have alluded earlier to the fact
that for gapped systems with a relatively small correlation
length---such as the two-leg ladder materials studied
here---incorporating as much as possible of the short-distance
structure of the system into the definition of the basis is clearly
advantageous~\cite{piek98,dmrmg97}. The three angular-momentum coupled
bases considered here are defined as the eigenstates of the
$2\!\times\!L$ Heisenberg ladder, with $L\!=\!1$, 2, and 4; for
$L\!=\!1$ and 2 these are the rung and plaquette basis, respectively.
For simplicity---and because of their intrinsic interest---we limit
ourselves to the physically relevant case of isotropic
$(J_{\perp}=J_{\parallel}\equiv 1)$ ladders.

In Table~\ref{tablethree} we list results for the ground-state
energy-per-site and for the single-triplet gap in various
approximations; the next to last column contains the
exact---basis-independent---results obtained from the DMRG calculation
of Ref.~\cite{whit94}, while the last column lists the spin-singlet
condensate fraction. This last column makes evident that the
``purity'' of the ground-state increases as one incorporates more and
more of the short-range structure of the system. However, as severe
truncations must be made, the overall picture does not necessarily
improve with the size of the basis. The second column contains the
zeroth-order results. These results are obtained from the mere
definition of the basis and are independent of any approximation.  We
observe that for these gapped systems incorporating as much as
possible of the short-range dynamics into the definition of the basis
is, indeed, advantageous. For example, for the $2\!\times\!4$ basis
the zeroth-order result for the energy is already within seven percent
of the exact answer. For the gap---a much more sensitive
quantity---the result still differs from the exact answer by 50
percent. Yet, this is significantly better than the factor-of-two
discrepancy obtained in the rung and plaquette bases.  The third
column lists the mean-field results. We observe that for the
energy-per-site the reduction relative to the zeroth-order values is
in good agreement with the second-order estimate from
Eq.~(\ref{perte}). That is,
\begin{equation}
     [\Delta E_{0}/{\rm site}]_{\rm{\scriptscriptstyle MF}} \simeq
     \cases{-0.18750 & for the $2\times 1$ basis; \cr
            -0.04167 & for the $2\times 2$ basis; \cr
            -0.01054 & for the $2\times 4$ basis. \cr}
 \label{mfcorrection}
\end{equation}
It is interesting to note that although the zeroth-order estimate
for the energy-per-site varies substantially among the three bases 
(up to 30 percent) the mean-field approximation brings all of them 
to within one percent of each other---and to within approximately
five percent of the exact answer. In the fourth column we have 
included the first-order correction coming from $H_{2}$ 
(see Sec.\ref{sec:pertur}): 
\begin{equation}
 [\Delta E_{0}/{\rm site}]_{\scriptscriptstyle H_{2}} =
 \cases{-0.04564 & for the $2\times 1$ basis; \cr
        -0.00386 & for the $2\times 2$ basis; \cr
        -0.00050 & for the $2\times 4$ basis. \cr}
 \label{mftcorrection}
\end{equation}
In all cases the correction to the energy-per-site is small relative
to the mean-field contribution, and justifies---albeit 
{\it a posteriori}---treating $H_{2}$ perturbatively.

The situation changes dramatically, however, upon examination of the 
singlet-triplet gap. In contrast to the energy-per-site which has a 
lowest-order contribution proportional to $\lambda^2$, the
mean-field contribution to the spin gap starts with the term 
$-2\lambda|\alpha|$. Although, for 
$\lambda\!=\!1$, the correction from the higher-order terms is not
negligible, this linear term dominates. Note that while the
mean-field correction to the spin gap is close to 40 percent in the
$2\!\times\!4$ basis, the corresponding correction to the
energy-per-site is a meager 2 percent. Moreover, it is now no longer
justifiable---specifically in the case of the rung basis---to treat
$H_{2}$ perturbatively. Indeed, the almost-perfect agreement between
the rung-basis mean-field and the exact answer is lost. While, on its 
own, this result does not invalidate the mean-field approach, a 
careful re-assessment of the role of $H_{2}$ is clearly necessary. 

The impact of $H_{2}$ on the gap, however, seems to be more 
moderate for the larger bases. For example, there is only a ten and
three percent correction to the gap in the plaquette and
$2\!\times\!4$ basis, respectively. This seems to indicate that the
whole approach, namely, mean-field plus $H_{2}$, is under control for
these larger bases. Yet the agreement with the exact result in the 
particular case of the $2\!\times\!4$ basis is poor (of the order of 
15 percent). We attribute this behavior to the severe truncation of 
the basis; recall that in the $2\!\times\!4$ basis only four out of 
a total of 256 states have been retained. 

\begin{figure}
\centerline{
  \psfig{figure=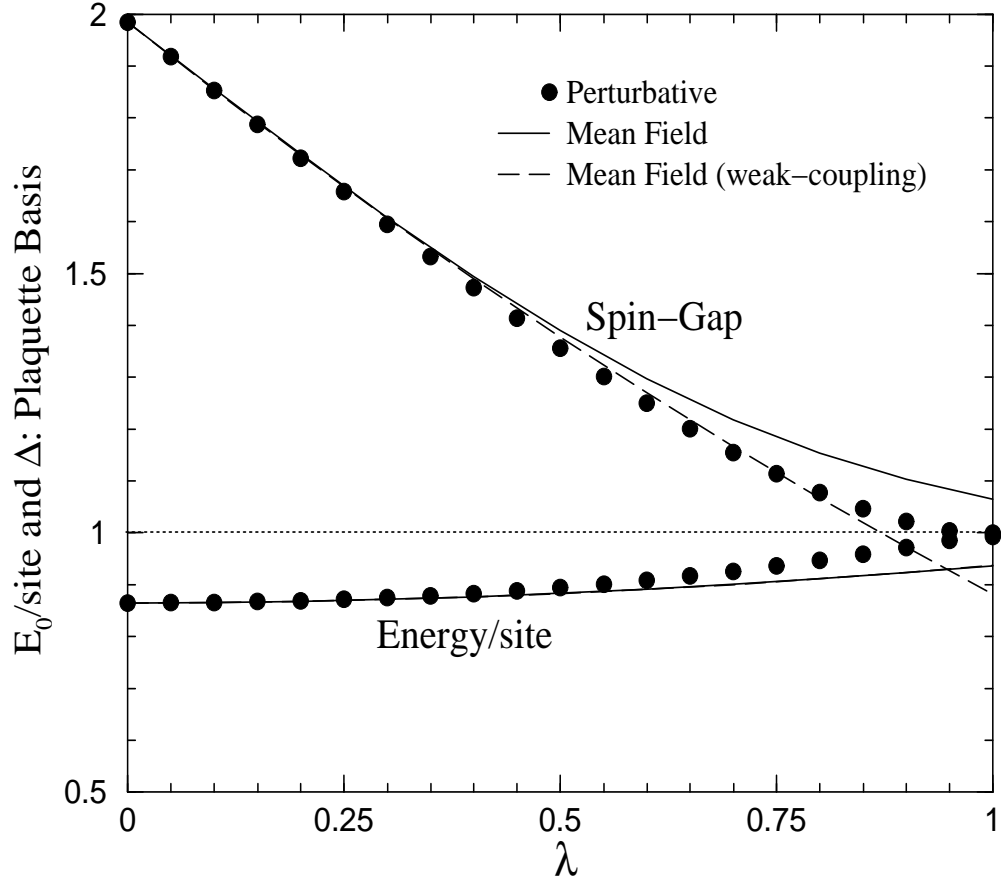,height=4.8in,width=5.2in,angle=0}}
 \vskip 0.1in
 \caption{Energy-per-site and singlet-triplet gap as a function 
	  of $\lambda$ in the plaquette basis. Quantities plotted
          are ratios to their exact values at the isotropic 
          ($\lambda=1$) point. The filled circles are from high-order
	  perturbative calculations~\protect\cite{piek98a}.} 
 \label{figsix}
\end{figure}

Overall, it seems that the plaquette basis offers the best
compromise. While there are non trivial perturbative corrections to
the gap from $H_{2}$, these are moderate---unlike the corrections
computed in the rung basis. Moreover, while we do truncate the basis,
the truncation errors are not as severe as in the $2\!\times\!4$
basis.  This is reflected in the ``bracketing'' of the exact result
for the gap between the mean-field and the mean-field plus $H_{2}$
estimates.  Further, because of the ``small'' expansion parameter in
the plaquette basis $[(\lambda\alpha/\Delta_{0})^{2}=\lambda^{2}/9]$
most of the mean-field behavior can be reproduced with terms up to
order $\lambda^2$ in the perturbative expansion. This is illustrated
in Fig.~\ref{figsix} where we have plotted the ground-state
energy-per-site and the spin gap as a function of $\lambda$. The
filled circles, essentially an exact calculation, are from a previous
perturbative estimate~\cite{piek98a}. Note that the $\lambda\!\ne\!1$
case represents a pair of dimerized chains connected to each other by
transverse bonds of strength equal to one of the two intra-chain
couplings. To the best of our knowledge, a physical realization of
this ``dimerized ladder'' is yet to be found. Yet, as it is well
known, the isotropic limit of these ladders has been the object of
intense scrutiny~\cite{dago96}.

\begin{figure}
\centerline{
  \psfig{figure=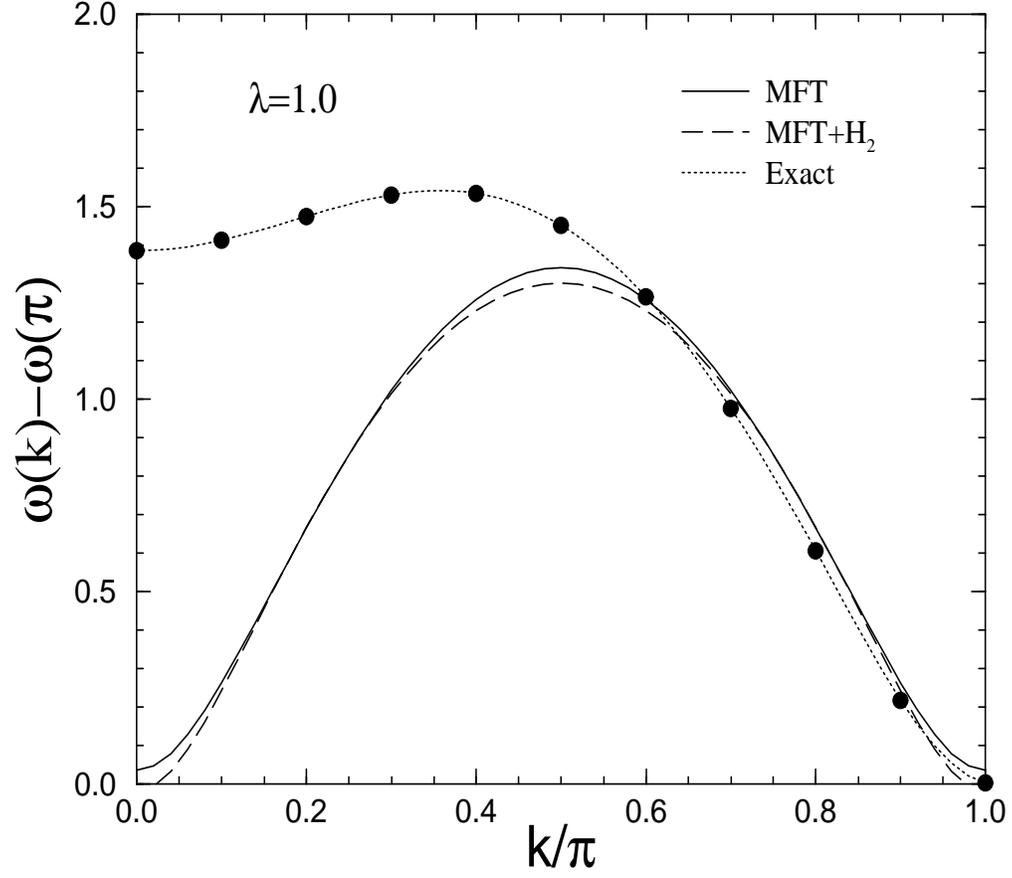,height=4.8in,width=5.2in,angle=0}}
 \vskip 0.1in
 \caption{Spin-triplet dispersion band relative to the band minimum 
	  in the plaquette basis as a function of momentum. The 
	  filled symbols represent exact results\protect\cite{barnes94}.}
 \label{figseven}
\end{figure}

\begin{figure}
\centerline{
  \psfig{figure=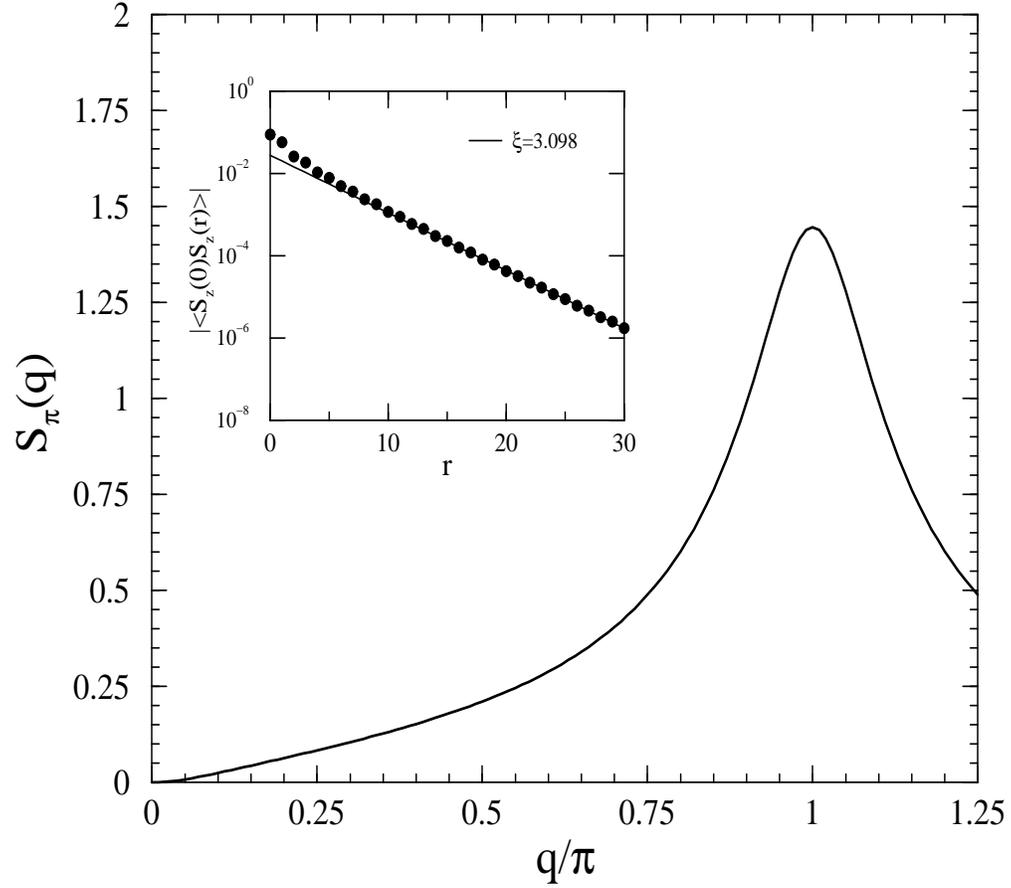,height=4.8in,width=5.2in,angle=0}}
 \vskip 0.1in
 \caption{Static structure factors $S_{\pi}$ in the plaquette
          basis as a function of $q$ at the isotropic point; 
          the structure factor $S_{0}$ vanishes in the (low-energy) 
          plaquette basis. The inset shows a logarithmic plot of 
	  its Fourier transform; the indicated straight-line fit yields
	  a value of $\xi\!=\!3.098$ for the spin-spin correlation 
	  length.} 
 \label{figeight}
\end{figure}

We conclude this section with a discussion of plaquette-basis results
for the dynamic spin response $S_{\pi}(q,\omega)$; recall that in a
mean-field approximation with the low-energy (truncated) plaquette
basis the $S_{0}(q,\omega)$ response is identically zero.  As the
strength of the $S_{\pi}$ response is concentrated in a single
excitation, it is sufficient to present results for the one-magnon
dispersion band and for the static structure factor in order to fully
characterize its behavior. Thus, in Fig.~\ref{figseven} we display the
dispersion relation for the one-magnon band at the isotropic value.
The solid and dashed lines (practically indistinguishable in the
figure) represent the mean-field and the mean-field plus $H_{2}$
results, respectively. The filled circles were computed from a a fit
to Lanczos data~\cite{barnes94}. We observe that the plaquette results
describe accurately the high-momentum/low-energy part of the band, but
deteriorate rapidly as one moves to the low-momentum/high-energy
region. This behavior is to be expected, as the discarded
(high-energy) states in the plaquette basis account for most of the
high-energy properties of the systems.  In the near future we are
planning to incorporate additional high-energy states into the
mean-field model in order to improve the description of the
high-energy properties of the system. Fortunately, as is evident in
Fig.~\ref{figeight} [see also Fig.~\ref{figfive}], the discarded
high-energy states are of little consequence to the dynamic spin
response, which has most of its strength concentrated at low
excitation energy. Moreover, this is the region that controls the
long-range behavior of the spin-spin correlation function, which is
shown in the inset on Fig.~\ref{figeight}. According to 
Eq.~(\ref{s0piplaq}), the plaquette-basis spin-spin correlation 
function per site is then given by
\begin{equation}
  \langle\Phi_{0}|S_{z}(0)S_{z}(r)|\Phi_{0}\rangle =
  \frac{\bar{s}^{2}}{6} \int_{0}^{2\pi}\frac{dq}{2\pi}\ 
  {e^{iqr}}\ \frac{\sin^2(q/2)}{\sqrt{1+d\cos(2q)}} \;.
 \label{spspin2p}
\end{equation}
A steepest-descent evaluation of the integral on the rhs of this 
equation yields, in analogy with Eq.~(\ref{spspin3}) for the rung basis,
\begin{equation}
  \langle\Phi_{0}|S_{z}(0)S_{z}(r)|\Phi_{0}\rangle \simeq
  e^{-r/\xi}\;; \quad
  \xi=2/\ln\left[-d/(1-\sqrt{1-d^2})\right] 
 \label{spspin3p}
\end{equation}
where we must recall that, in the plaquette basis, $d\!<\!0$, 
while $d$ is positive in the rung basis. A fit to the full result 
for the correlation function (see Fig.~(\ref{figeight}))
gives a value of $\xi\!=\!3.098$ for the correlation length, which is
within seven percent of the approximate analytic value of
\begin{equation}
   \xi\approx \frac{2}{\ln\left[-d/(1-\sqrt{1-d^2})\right]}  
   \smash{\mathop{\longrightarrow}\limits_{d=-0.843}}=3.327 .
 \label{xiplaq}
\end{equation}
These results compare very well---at the three-percent level---with
the exact value of $\xi\!=\!3.19$ obtained from the DMRG analysis of
Ref.~\cite{whit94}. Our results for correlation lengths using both 
rung and plaquette bases are summarized in Table~\ref{tablefour}.

\section{Conclusions}
\label{sec:conclusions}
We have computed magnetic properties of Heisenberg spin-$1/2$ ladders
using an extension of the mean-field approximation of Gopalan, Rice, 
and Sigrist~\cite{gopa94} which we have implemented using a variety of 
angular-momentum coupled bases. In the original version of the 
model---which employed the rung basis exclusively---the Heisenberg 
Hamiltonian was written in terms of
spin-singlet and spin-triplet operators. Because of the rotational
symmetry of the Hamiltonian, the full theory is specified in terms of
only four parameters; these are the rung-singlet and rung-triplet
energies, and two off-diagonal matrix elements. In the limit in which
the ratio $(\lambda)$ of the coupling of the spins along the chains
of the ladder relative to the coupling along the rungs is negligible, 
the ground-state wavefunction becomes a direct product of
spin-singlets on every rung of the ladder. The first spin excitation 
is reached by breaking a spin-singlet along any of the rungs of the 
ladder. This weak-coupling limit of the theory is the motivation 
behind the mean-field approximation of Ref.~\cite{gopa94}. Hence,
in this approximation, one assumes that the spin-singlet
bosons of zero linear momentum are condensed. This enables one to 
separate the Heisenberg Hamiltonian into a part that is quadratic 
in the spin-triplet operators ${\bf t}$---the mean-field 
Hamiltonian---and a part $(H_{2})$ that remains quartic in 
${\bf t}$. The mean-field Hamiltonian is then brought to a 
diagonal form by means of a Bogoliubov transformation while the 
quartic part must be treated by some other means.

As the mean-field Hamiltonian is determined by only a small number of
parameters, its generalization to larger bases is straightforward. The
physical motivation behind the change of basis is the small value (of
the order of three lattice sites) of the spin-spin correlation
length. Thus, the main advantage of the larger bases is that they can
incorporate exactly---by their mere definition---part of the important
short-range dynamics of these gapped systems.

Our first calculations were carried out in the rung basis, as in
Ref.~\cite{gopa94}. We implemented the mean-field procedure through a
Bogoliubov transformation, which although standard, was effected here
via a diagonalization of a suitable RPA-like matrix. One of the
virtues of such an approach is that the generalization to larger
vector spaces becomes simple. This is of special interest to us, as
our future mean-field work will incorporate additional high-energy
states of the plaquette basis. Our rung-basis results differ from
those presented by the authors of Ref.~\cite{gopa94} due to what we
believe was an oversight on their part. Our results indicate that,
contrary to their first reports, their mean-field approach works well
at a quantitative level. Indeed, we find that the results for the
energy-per-site and for the singlet-triplet gap agree with exact
calculations to within six percent all through the
$0\!\le\!\lambda\!\le\!1$ range (note, however, that the value for the
spin-spin correlation length at the isotropic point was heavily
underestimated: $\xi\!=\!2.211\!<\!3.19$). In particular, the spin gap
differs from the exact answer by less than one percent at the
isotropic point. Unfortunately, this nice agreement was lost once the
contribution from $H_{2}$ was estimated perturbatively. Incorporating
the effects from $H_{2}$ in a reliable manner should constitute an
important goal for future work.

For the larger $2\times 2$ (plaquette) and $2\times 4$ bases
considered here an additional approximation was required. We assumed
that the important physics of the spin-ladder materials is dominated
by the lowest singlet and triplet states in the corresponding bases;
note that it is only in the rung basis that the singlet and triplet
states are unique. We found that while rung- and plaquette-basis
results are of similar quality for most ladder observables, only the
latter ones were stable upon including the perturbative contribution 
from $H_{2}$. This is a direct consequence of the size of the 
elementary block defining the basis. As the larger bases incorporate 
more of the short-range dynamics of the system, the quartic piece of the
Hamiltonian ($H_{2}$) decreases in importance as the size of the
elementary block increases. Since our general approach is easily
extended to any angular-momentum coupled basis, mean-field studies
were also carried out utilizing a $2\times 4$ basis. The results
obtained with this basis did not improve---and in some cases
worsen---the agreement relative to the plaquette basis. We have
traced this behavior to the severe truncations made in the $2\times 4$
basis; only four out of a total of 256 states were retained. Thus,
we conclude that the plaquette basis offers the best compromise 
between building enough short-range physics into the definition of
the basis, while avoiding too severe a truncation. Indeed, 
plaquette-basis mean-field results for the energy-per-site, the spin
gap, and the spin-spin correlation function are within five percent
of the quoted DMRG values~\cite{whit94}.

In the future, we plan to complete a number of extensions of the
mean-field treatment. For example, we have truncated the spectrum of
spin triplet states in the plaquette basis retaining only those of
lowest energy. Although we have observed that this is a good
approximation for describing the low-energy dynamics, it is not
adequate to describe the high-energy properties. This requires an
extension of the usual Bogoliubov technique for bringing quadratic
Hamiltonians involving a single species of field operator into
diagonal form. We have already made considerable progress in this
direction and we are planning to publish our results in the near
future. Alternatively, one could still retain the truncated basis but 
account for the truncation errors through a suitable renormalization 
of the matrix elements. Indeed, one such technique---CORE, or the 
contractor renormalization group method~\cite{morwei94}---has been used
successfully by us in computing static and dynamic properties of up to
$2\!\times\!16$ ladders~\cite{piek98}. Such an approach, to be used in
the near future, could also be valuable in the context of the
mean-field theory. We are also planning to develop a version of the
mean-field theory for ladders which treats the quartic terms of the
Hamiltonian self-consistently, rather than just perturbatively as it
was done here. Finally---and of special relevance to the high-$T_{c}$
materials---we are planning to use the mean-field theory in the
various angular-momentum coupled bases to study the physics of
lightly-doped ladders~\cite{sig94}.

\appendix
\section*{Bogoliubov Transformation}
\label{sec:appendix}
In this appendix we carry out the Bogoliubov transformation
via a diagonalization of a random-phase-approximation (RPA) matrix. 
In order to do so we write the---operator part---of the mean-field 
Hamiltonian of Eq.~(\ref{kmft}) in matrix form. That is,
\begin{equation}
 \hat{K}_{\rm MF}(\bar{s}^2,\mu)  \equiv 
     \frac{1}{2}\sum_{km}
     {\cal T}^{\dagger}_{m}(k)\,\Omega(k)\,{\cal T}_{m}(k) \;.
 \label{kmfta}
\end{equation}
Although we illustrate the method using the $2 \times 2$ matrix of 
Eq.~(\ref{todef}) the method can be easily generalized to matrices 
of arbitrary dimension. 

Given that the matrix $\Omega(k)$ is symmetric, it can be diagonalized
via a similarity transformation. The problem with such an approach, 
however, is that the resulting orthogonal matrix generates a new set 
of creation and annihilation operators that are not consistent with
each other. Instead, what is required is to write Eq.~(\ref{kmfta}) 
in terms of a matrix having an RPA-like structure, i.e.,
\begin{equation}
 \hat{K}_{\rm MF}(\bar{s}^2,\mu)  = 
     \frac{1}{2}\sum_{km}
     {\cal T}^{\dagger}_{m}(k)\,g\,\Omega_{\rm RPA}(k)\,
     {\cal T}_{m}(k) \;,
 \label{kmftrpa}
\end{equation}
where we have introduced the ``metric'' $g\equiv{\rm diag\{1,-1\}}$ 
and have defined $\Omega_{\rm RPA}\equiv g\Omega$. We now list,
without proof, some useful properties of RPA matrices:
\begin{enumerate}
 \item{Eigenvalues of $\Omega_{\rm RPA}$ come in pairs; 
       if $\omega$ is an eigenvalue of $\Omega_{\rm RPA}$ 
       so is $-\omega$}.
 \item{Eigenvectors of $\Omega_{\rm RPA}$ corresponding to different
       eigenvalues are orthonormal with respect to the metric $g$:
       $\langle\omega'|g|\omega\rangle=
        {\rm sgn}(\omega)\,\delta_{\omega'\omega}$.}
 \item{There exists a ``similarity'' transformation ${\cal S}$---with its
       columns being the eigenvectors of $\Omega_{\rm RPA}$---that 
       brings $\Omega_{\rm RPA}$ into a diagonal form:
       ${\cal S}^{-1}\,\Omega_{\rm RPA}\,{\cal S}=\!
       {\rm diag}\{\omega,-\omega\}\,,$ where
       ${\cal S}^{-1}\!=g\,{\cal S}^{T}g$}. 
\end{enumerate}
Using these properties, and defining a new set of creation and 
annihilation operators through  
$\Gamma_{m} \equiv {\cal S}^{-1}{\cal T}_{m}$, the mean-field Hamiltonian 
becomes explicitly diagonal. That is,
\begin{equation}
 \hat{K}_{\rm MF}(\bar{s}^2,\mu)  = 
     \frac{1}{2}\sum_{km}
     \Gamma_{m}^{\dagger}(k)
     \pmatrix{\omega(k) &  0          \cr
                 0      &  \omega(k) \cr}
     \Gamma_{m}(k) = \frac{3}{2}\sum_{k}\omega(k)
     + \sum_{km}\omega(k)\gamma_{m}^{\dagger}(k)\gamma_{m}(k) \;.
 \label{kmftadiag}
\end{equation}

\acknowledgments
This work was supported by the DOE under
Contracts Nos. DE-FC05-85ER250000, DE-FG05-92ER40750 and
DE-FG03-93ER40774.


\vfill\eject
\vfill\eject
\vfill\eject
\vfill\eject
\vfill\eject
\vfill\eject
\vfill\eject
\vfill\eject
%
\mediumtext
 \begin{table}
  \caption{Unperturbed singlet/triplet energies and the 
	   two independent matrix elements in 
           various angular-momentum-coupled bases.}
   \begin{tabular}{lcccc}
     Basis     & $\epsilon_{0}$  & $\epsilon_{1}$
               & $\alpha\equiv \langle 011||V(1,2)||101\rangle$ 
               & $\beta \equiv \langle 110||V(1,2)||110\rangle$ \\
   \tableline
   $2 \times 1$  & $-3/4$  &  $+1/4$  &  $+1/2$  &  $-1  $ \\   
   $2 \times 2$  & $-2  $  &  $-1  $  &  $-1/3$  &  $-1/4$ \\
   $2 \times 4$  & $-4.29307$ & $-3.52286$ & 
                   $-0.20805$ & $-0.08625$ \\
   \end{tabular}
  \label{tableone}
 \end{table}
\mediumtext
 \begin{table}
  \caption{Results for the ground-state energy-per-site and for
           the singlet-triplet gap (displayed inside brackets)
	   for a few values of $\lambda$ in the Rung basis.}
   \begin{tabular}{cccc}
   $\lambda$ & Ref.~\cite{gopa94} & Present work & Exact    \\
   \tableline
      $0.25$  &  $-0.37908\;[0.72846]$  &  $-0.38683\;[0.76699]$  
              &  $-0.38804\;[0.78438]$                         \\
      $0.50$  &  $-0.39392\;[0.40411]$  &  $-0.42237\;[0.59685]$    
              &  $-0.42911\;[0.64065]$                         \\
      $1.00$  &  $-0.47531\;[0.10291]$  &  $-0.54285\;[0.50133]$  
              &  $-0.57804\;[0.50400]$                         \\
   \end{tabular}
  \label{tabletwo}
 \end{table}
\mediumtext
 \begin{table}
  \caption{Results for the ground-state energy-per-site and for
           the singlet-triplet gap (displayed inside brackets)  
	   for various angular-momentum coupled bases. Listed 
	   are zeroth-order, mean-field, mean-field plus $H_{2}$, 
	   and exact results (see text for details). The exact 
	   values were extracted from DMRG calculations. The last
           column lists the spin-singlet condensate fraction.}
   \begin{tabular}{cccccc}
     Basis & Zeroth-order & MFT & MFT+$H_{2}$ & Exact & $\bar{s}^2$ \\
   \tableline
    $2 \times 1$ & $-0.37500\;[1.00000]$ & $-0.54285\;[0.50133]$ 
                 & $-0.58849\;[0.13204]$ & $-0.57804\;[0.50400]$  
                 & $ 0.78129$ \\    
    $2 \times 2$ & $-0.50000\;[1.00000]$ & $-0.54112\;[0.53689]$ 
                 & $-0.54498\;[0.48154]$ & $-0.57804\;[0.50400]$  
                 & $ 0.85826$ \\  
    $2 \times 4$ & $-0.53663\;[0.77021]$ & $-0.54724\;[0.44566]$ 
                 & $-0.54774\;[0.43216]$ & $-0.57804\;[0.50400]$  
                 & $ 0.89628$ \\  
   \tableline
   \end{tabular}
  \label{tablethree}
 \end{table}
\mediumtext
 \begin{table}
  \caption{Results for spin spin-spin correlation lengths for the 
	   rung and plaquette bases. The ``Fit'' results are obtained
	   by straight-line fits to semi-log plots of the correlation 
	   function as determined by numerical evaluation of the integral 
	   appearing in Eqn.(\protect\ref{spspin}). The ``Analytic'' 
	   results are obtained using Eqns.(\protect\ref{xirung}) and
	   (\protect\ref{xiplaq}) for rung and plaquette bases, 
	   respectively. The ``Exact'' value is from the DMRG 
	   calculation of Ref.~\protect\cite{whit94}.}
   \begin{tabular}{cccc}
     Basis & Fit & Analytic & Exact \\
   \tableline
    $2 \times 1$ & $2.211$ & $2.323$ & $3.19$ \\    
    $2 \times 2$ & $3.098$ & $3.327$ & $3.19$ \\
   \tableline
   \end{tabular}
  \label{tablefour}
 \end{table}
\end{document}